\documentclass[aps,twocolumn,prl,superscriptaddress,amsmath,amssymb,showpacs]{revtex4-1}
\usepackage{graphicx}
\usepackage{amsmath}
\usepackage{comment}
\usepackage{bm}
\usepackage{color}

\renewcommand{\thesection}{\Roman{section}}

\begin{document}

\title{Ripple state 
in the frustrated honeycomb-lattice antiferromagnet}

\author{Tokuro Shimokawa} 
\email{tokuro.shimokawa@oist.jp}
\affiliation{Okinawa Institute of Science and Technology Graduate University, Onna, Okinawa, 904-0495, Japan}
\author{Hikaru Kawamura}
\affiliation{Department of Earth and Space Science, Graduate School of Science, Osaka University, Toyonaka, Osaka 560-0043, Japan}

\date{\today}

\begin{abstract}
We discover a new type of multiple-$q$ state, ``ripple state", in a frustrated honeycomb-lattice Heisenberg antiferromagnet under magnetic fields. The ground state has an infinite ring-like degeneracy in the wavevector space, exhibiting a cooperative paramagnetic state, ``ring-liquid'' state. We elucidate that the system exhibits the ripple state as a new low-temperature thermodynamic phase via a second-order phase transition from the ring-liquid state, keeping the ring-like spin structure factor. The spin texture in real space looks like a ``water ripple" and can induce a giant electric polarization vortex. Possible relationship to the honeycomb-lattice compound, ${\rm Bi_{3}Mn_{4}O_{12}(NO_{3})}$, is discussed.

\end{abstract}

\maketitle

\textcolor{black}{In the past decade, a honeycomb-lattice system has attracted much attention in condensed matter physics, in the context of, {\it e.g.\/}, graphene~\cite{Geim}  or the so-called Kitaev spin liquid~\cite{Kitaev}.}
Even before the advent of the Kitaev model, the $J_1$-$J_2$ classical antiferromagnetic (AF) Heisenberg model on the honeycomb-lattice  with the competing nearest-neighbor (NN) and next-nearest-neighbor (NNN) interactions, $J_1$ and $J_2$ shown in Fig.~\ref{Fig1}(a), was known as a possible candidate of a spin liquid state~\cite{Anderson}. Although the honeycomb lattice is bipartite and can accommodate the conventional two-sublattice AF order, the frustration effect due to the competition between the NN and NNN interactions leads to highly degenerate ground states \cite{Rastelli, Katsura}, and the associated fluctuations tend to suppress the spin ordering. In the ground state of the classical system under zero magnetic field, an infinitely degenerate manifold consisting of a set of spiral states characterized by generally incommensurate wavevectors ${\bf q}$, which form the ring-like pattern surrounding the AF point in the reciprocal space, appears when $J_2$ is moderately strong, $1/6<J_2/J_1<0.5$. Typical ground-state ring-like degenerate lines are shown in Fig.~\ref{Fig1}(b) for several $J_2$/$J_1$-values.

\begin{figure*}
  \includegraphics[bb=0 50 792 550, width=17.5cm,angle=0]{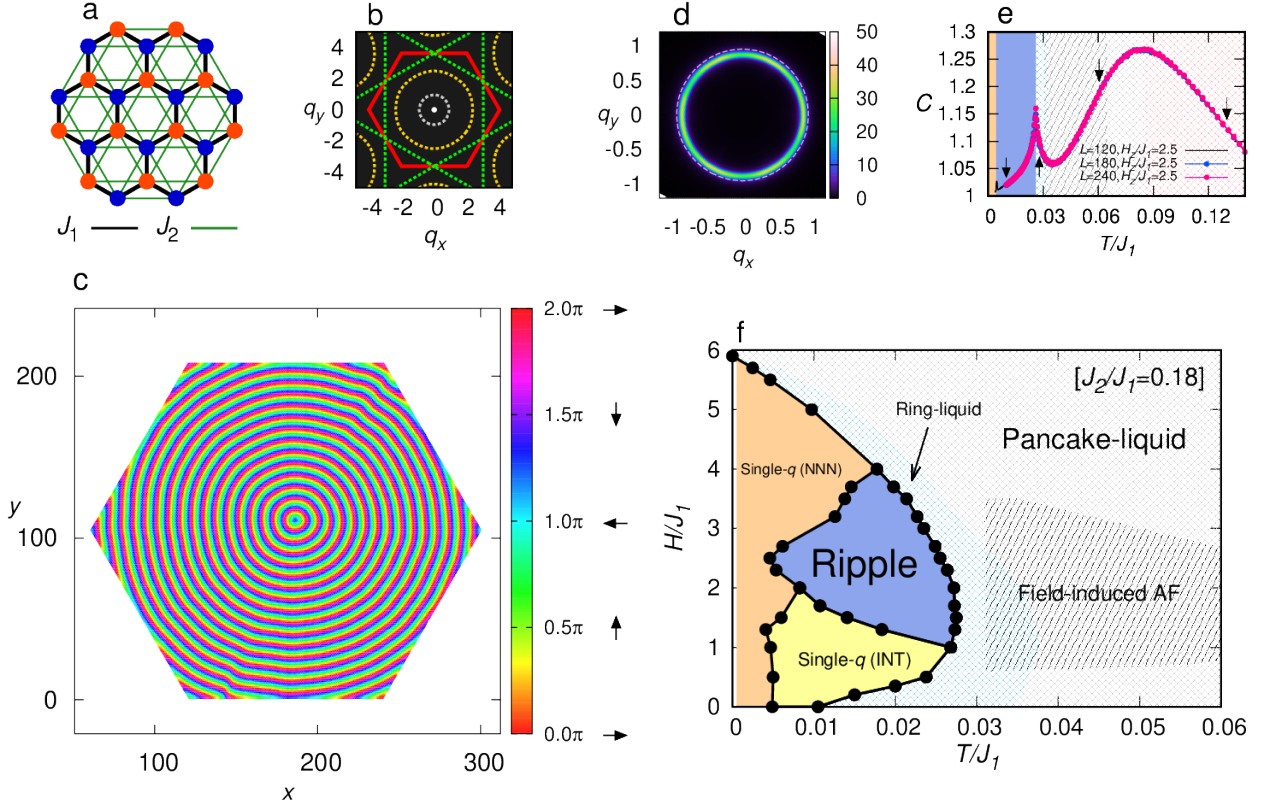}
 \caption{
 (a) The $J_1$-$J_2$ classical honeycomb-lattice Heisenberg antiferromagnet with the nearest-neighbor (black line) interaction $J_1$, and the next-nearest-neighbor (green line) interaction $J_2$. The lattice consists of two interpenetrating triangular sublattices, A (red site) and B (blue site). The lattice constant of the  triangular sublattice, equal to the next-nearest-neighbor distance of the honeycomb lattice, is taken to be the length unit.
(b) The ground-state manifold lines of the model in the sublattice wavevector space for $J_2/J_1=0.18$ (gray), $0.3$ (yellow), and $0.5$ (green), where a white point at the origin ${\bf q}=0$ is the AF point. Red solid hexagon indicates the first Brillouin zone of the triangular sublattice.  
(c,d) The real-space sublattice $xy$-spin configuration (c), and the corresponding sublattice static spin structure factor (d), in the ripple state at $J_2/J_1=0.18$. The temperature and the magnetic field are $T/J_1$=0.01\textcolor{black}{0294} and $H/J_1=2.5$, the system size being $L$=240 ($N$=86400 spins) under open BC. In (c), the color means the direction of the $xy$-spin, while in (d) the dashed gray line displays the ground-state degeneracy line for $J_2/J_1$=0.18.
(e) The temperature dependence of the specific heat per spin for $J_2/J_1=0.18$ and $H/J_1=2.5$, covering from the higher-temperature paramagnetic to the lower-temperature single-$q$ (NNN) states. The lattice sizes are $L=120$, 180 and 240 under open BC. Each arrow indicates the point at which the spin structure factor is shown in Fig.~\ref{Fig1}(d), and in Fig.~S1(b), (d) and (f) of Supplemental Material.
(f) $T$-$H$ phase diagram for $J_2/J_1=0.18$ obtained by MC simulations.
}
 \label{Fig1}
\end{figure*}

Interest in this $J_1$-$J_2$ honeycomb antiferromagnet has been accelerated by the recent experiments on the $S$=3/2 bilayer honeycomb-lattice Heisenberg AF ${\rm Bi_3 Mn_4 O_{12} (NO_3)} $\cite{Smirnova, Matsuda}, which exhibits a spin-liquid-like behavior down to a low-temperature of 0.4 K. Surprisingly, in spite of the short AF correlation length of a few lattice spacings in zero field, a field-induced antiferromagnetism was observed even for weak fields of 6 [T]. Okumura $et$ $al$ \cite{Okumura} theoretically pointed out that the origin of the observed spin-liquid-like behavior might be the ring-like degeneracy \textcolor{black}{specific} to the $J_1$-$J_2$ honeycomb model. While the model turned out to exhibit a single-$q$ spiral \textcolor{black}{state} at low enough temperatures for $J_2/J_1>1/6$ due to the order-by-disorder mechanism, the associated energy scale can be arbitrarily small near the AF phase boundary at $J_2\rightarrow \frac{1}{6}J_1$, giving rise to exotic spin-liquid states called ``ring-liquid" and ``pancake-liquid" states. The latter state indeed accompanies the field-induced antiferromagnetism\cite{Okumura}. The spin structure factor in the ring-liquid state exhibits a ring-like pattern surrounding the AF point in the wavevector space reflecting the ground-state degeneracy, while, in the pancake-liquid state, the center of the ring is `buried' in intensity, yielding a pancake-like pattern see also Fig.~S1(d) and (f) of Supplemental Material).

 In-field ordering of frustrated Heisenberg magnets has also attracted much recent interest, since exotic multiple-$q$ ordered states with nontrivial spin textures are often realized there \cite{Okubo,Seabra}. 
One interesting example might be the $J_1$-$J_3$ ($J_1$-$J_2$) \textcolor{black}{ triangular-lattice Heisenberg magnet}, which exhibits an intriguing triple-$q$ \textcolor{black}{state} corresponding to the chiral-symmetric skyrmion-lattice state even in the absence of the Dzyaloshinskii-Moriya interaction \cite{Okubo}.
\textcolor{black}{Recent studies have also revealed that the in-field ordering of the $J_1$-$J_2$ honeycomb Heisenberg antiferromagnet sustains a variety of exotic multiple-$q$ states for moderately large values of $J_2/J_1 \gtrsim 0.2$ \cite{Rosales, Shimokawa}. 
Yet, the multiple-$q$ states there consist of only a finite number of ${\bm q}$-vectors selected from the degenerate ring by breaking its continuous degeneracy via the order-by-disorder mechanism, and in that sense, are the standard ordered states. Furthermore,} the most interesting  $J_2/J_1$-region close to the AF phase boundary, where fluctuation effects combined with the ring-like infinite degeneracy are most pronounced, remain unexplored.

In this Letter, we report on our finding of a new frustration-induced multiple-$q$ state, a ``ripple state", appearing in the $J_1$-$J_2$ honeycomb Heisenberg model in the vicinity of the AF phase boundary. The spin texture in the ripple state, shown in Fig.~\ref{Fig1}(c), consists of an infinite number of the spirals running along every direction from a core. Hence, the spin configuration in real space looks like a ``water ripple", while the corresponding spin structure factor shown in Fig.~\ref{Fig1}(d) keeps the ring-like pattern as the ring-liquid state does. In contrast to the ring-liquid paramagnetic state, the ripple state is a thermodynamic ordered state accompanied with a spontaneous symmetry breaking, which emerges from the ring-liquid paramagnetic state via a second-order transition. This can be deduced, {\it e.g.\/}, from the sharp specific-heat anomaly at the onset of the ripple state, as shown in Fig.~\ref{Fig1}(e). Surprisingly, the mechanism of order by disorder does not work in this state, and all wavevectors on the degenerate ring equally contribute to its order.

\begin{figure*}[t]
  \includegraphics[bb=0 100 792 500,width=18.0cm,angle=0]{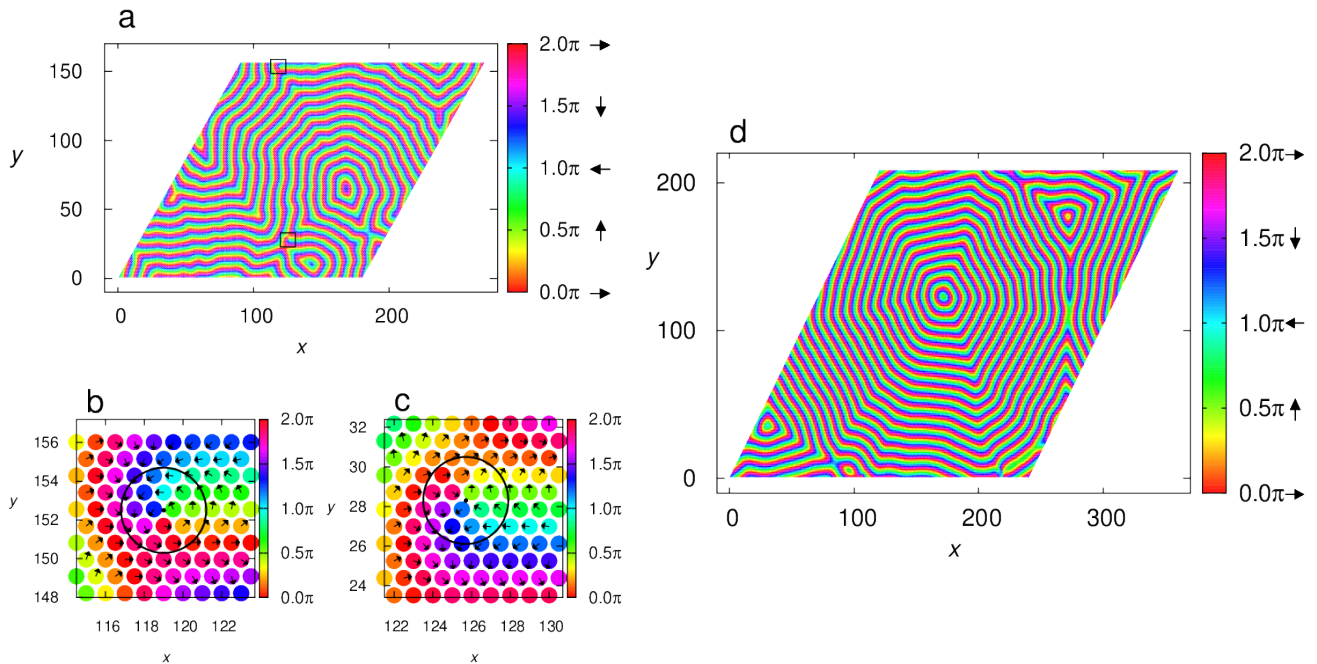}
 \caption{(a) The real-space sublattice $xy$-spin configuration in the ripple state at $J_2/J_1=0.18$, $T/J_1$=0.015 and $H/J_1$=2.5. The lattice is a diamond cluster of the size $L$=180 ($N$=64800 spins) under periodic BC.
(b-c) Focused views of the regions denoted by black squares in (a). Arrows indicates the directions of the $xy$-spins. A vortex (b) and an anti-vortex (c) patterns are detectable: see the circle in the figure.
(d), The real-space sublattice $xy$-spin configuration in the ripple state in the same condition as (a), for a larger diamond cluster of $L$=240 ($N$=115200 spins) under periodic BC.}
 \label{Fig3}
\end{figure*}

Our Hamiltonian is given by
\begin{eqnarray}
\mathcal{H}=J_1\sum_{\langle i,j \rangle }{\bf S}_i \cdot {\bf S}_j + J_2\sum_{\langle \langle i,j \rangle \rangle}{\bf S}_i \cdot {\bf S}_j - H \sum_{i} S_i^{z},
\label{hami}
\end{eqnarray}
where ${\bf S}_i=(S_i^x, S_i^y, S_i^z)$ is the classical Heisenberg spin of the fixed length $|{\bf S}_i|=1$, $H$ is the field intensity, $\sum_{\langle i,j \rangle}$ and $\sum_{\langle \langle i,j \rangle \rangle}$ mean the sum over the NN and NNN pairs, respectively, with $J_1>0$ and $J_2>0$. As a representative parameter, we focus here on the case of $J_2/J_1$=0.18 close to the AF phase boundary $J_2/J_1=1/6$.

  By extensive MC simulations, the temperature ($T$) - magnetic field ($H$) phase diagram of the model is constructed as given in  Fig.~\ref{Fig1}(f). In addition to the ring-liquid and pancake-liquid paramagnetic states, there appears the ripple state introduced above, together with the two types of single-$q$ spiral \textcolor{black}{states} described as (NNN) and (INT), the $q$-vector running along the NNN direction in the former and along the low-symmetric intermediate direction in the latter.

 As shown in Fig.~\ref{Fig1}(e), on decreasing $T$, the specific heat exhibits a rounded peak at $T/J_1 \simeq 0.084$ corresponding to a crossover from the high-$T$ paramagnetic state to the pancake-liquid state,  a minimum at $T/J_1\simeq 0.035$ corresponding to a crossover from the pancake-liquid state to the ring-liquid state, then a sharp peak at $T_c/J_1\simeq 0.026$ corresponding to a phase transition from the ring-liquid state to the ripple state, and finally a small but clear peak at $T_c'/J_1\simeq 0.0045$ corresponding to a phase transition from the ripple state to the single-$q$ spiral state. The running direction of the spiral is entropically determined by thermal fluctuations via the order-by-disordered mechanism examined by the low-temperature expansion \cite{Okumura, Shimokawa, Bergman}.

Now, we describe each phase in some details, focusing on the ripple state. The details of other states in Fig.~\ref{Fig1}(f) are shown in Supplemental Material.
We mainly concentrate on the spin $xy$-components since the $z$-component parallel with the applied field usually exhibits the ${\bf q}$=${\bf 0}$ component only. Furthermore, since there is always a phase difference of nearly $\pi$ between the spin $xy$-components on the sublattices A and B, we focus below on the spins on one particular sublattice, say, the sublattice A (red sites in Fig.~\ref{Fig1}(a)). Further details of the spin ordering on the two sublattices are given in Supplemental Material.

The real-space sublattice $xy$-configuration and the associated sublattice spin structure factor $S_{\perp}({\bf q})$ are shown in Fig.~{\ref{Fig1}}(c) and \ref{Fig1}(d) for the ripple state, and in Fig.~S1 for various other states. In the ripple state, the spins form a water-ripple-like pattern in real space, extending in all directions from the sample core almost isotropically, in sharp contrast to the single-$q$ spiral state. 
Reflecting the lack of any directional anisotropy of the order, the associated $S_{\perp}({\bf q})$ keeps a full isotropy of the degenerate ring shown in Fig.~{\ref{Fig1}(d)}, in spite of the occurrence of a phase transition. In spite of the low temperature $T/J_1\simeq 10^{-2}$, the mechanism of order-by-disorder does not work, while it eventually works at a further lower temperature below the transition temperature into the single-$q$ state, $T_c'/J_1 \simeq 0.0045$: refer to the small specific-heat anomaly in Fig.~\ref{Fig1}(e). 

 Although the ripple state is well compatible with open BC applied in \textcolor{black}{Figs.~\ref{Fig1}(c) and S5 (see also Supplemental Material),} one may wonder if what happens to this state if the standard periodic BC is applied. Notice that a single ripple is clearly incompatible with periodic BC. Then, we also investigate the ordering properties under periodic BC. The real-space $xy$-spin configuration of the ripple-state is shown for the $L=180$ diamond cluster in Fig.~\ref{Fig3}(a). In order to match the imposed periodic BC, ripples are now deformed with more than one ripple cores generated together with the interference between them. The pattern resembles the interference between two real water-ripples. This interference also results in the appearance of ``branching points'' shown in the boxed area of Fig.~\ref{Fig3}(a), which may be viewed as defects of the ripple \textcolor{black}{state}. For two such defect points, we show in Fig.~\ref{Fig3}(b) and (c) their real-space $xy$-spin configuration. These two defect points actually correspond to a vortex and an anti-vortex. We carefully investigate the appearance pattern of the (anti)vortex across $T_c$ for both open and periodic BCs, to confirm that these (anti)vortices are externally introduced by the applied periodic BC, and are not thermal objects related the Kosterlitz-Thouless(KT)-type vortex unbinding transition \cite{Berezinskii,Kosterlitz}.

 Given that more than one ripple is generated at least under periodic BC, we also carefully study the possible size effect to check the possibility that the ripple state actually consists of a periodic array of ripples with very long periodicity. In Fig.~\ref{Fig3}(d), we show the real-space $xy$-spin configurations for a larger system of size $L=240$ under periodic BC. A ripple pattern very much similar to Fig.~\ref{Fig3}(a) is observed, {\it i.e.\/}, while more than one ripples are again generated, the size of a ripple is enlarged just as the system size $L$, demonstrating that the observed ripple deformation is the effect forced by periodic BC. \textcolor{black}{Implication of these observation is that, even in the thermodynamic limit, there will be only one huge ripple under open BC, or only a few huge ripples when forced by the particular BC, {\it e.g.\/}, periodic BC.} In particular, no super-lattice structure exists in the spatial arrangement of the ripple. This is in sharp contrast to the skyrmion states which are stabilized almost always in the form of a crystal \cite{Muhlbauer,Bogdanov}. 

 Even so, the translational symmetry is spontaneously broken in the ripple state, though in an algebraic manner (recall that the model is two-dimensional (2D) and the relevant symmetry is continuous). To see this, we examine the size and time (MC steps) dependence of the real-space spin pattern, and some of the result is shown in Supplemental Material. We apply periodic BC to prevent the boundary serving as a pinning center to the ripple motion. As can be seen from Fig.~S4, the spin pattern is locked during a long MC time of $2\times 10^5$ steps, suggesting the spontaneous breaking of both the translation and the $U(1)$ spin-rotation symmetries.

 To further clarify the nature of the phase transition and the associated symmetry-breaking pattern, we examine the finite-size effects on physical quantities around the paramagnetic-to-ripple transition temperature $T_c$. Fig.~\ref{Fig4}(a) exhibits the temperature and size dependence of the specific heat across $T_c$, a magnified view of Fig.~\ref{Fig1}(e). The peak height grows with $L$, suggesting the divergent anomaly occurring at $T_c$. Fig.~\ref{Fig4}(b) exhibits the temperature and size dependence of the lattice threefold-rotation-symmetry ($C_3$) breaking parameter $m_3$ (the definition is given in Supplemental Material) at a field $H/J_1=2.5$ associated with the ripple \textcolor{black}{state}, and at $H/J_1=0.5$ associated with the single-$q$ spiral \textcolor{black}{state}. One can confirm that, although $C_3$ is spontaneously broken in the spiral state, it is unbroken in the ripple state.

 Thus, in the ripple state, while the translation and the $U(1)$ spin-rotation symmetries are spontaneously broken, the threefold lattice-rotation $C_3$ symmetry is unbroken. Since the $U(1)$-symmetry breaking in 2D should be of the KT type not accompanying a pronounced specific-heat anomaly, the origin of the observed divergent-like anomaly requires the other type of symmetry breaking. In fact, the model retains a $Z_2$ symmetry associated with the spin-reflection in spin space (not in real space) with respect to an arbitrary plane including the magnetic-field ($z$) axis, say, $(S_i^x,S_i^y,S_i^z) \rightarrow (S_i^x,-S_i^y,S_i^z)$. Such a $Z_2$-symmetry breaking is expected to accompany the specific-heat divergence similar to the one of the 2D Ising model.

 A convenient quantity to probe such a $Z_2$-symmetry breaking might be the {\it chirality}. We compute the scalar chirality $\chi$ defined for three neighboring spins on each elementary triangle to examine its real-space pattern in the ripple state. Indeed, the result indicates that the $Z_2$-symmetry is spontaneously broken in the ripple state. The details are given in Supplemental Material. We also construct a phenomenological equation reproducing the observed ripple spin texture. The details are also given in Supplemental Material.

\begin{figure}[t]
  \includegraphics[width=9.0cm,angle=0]{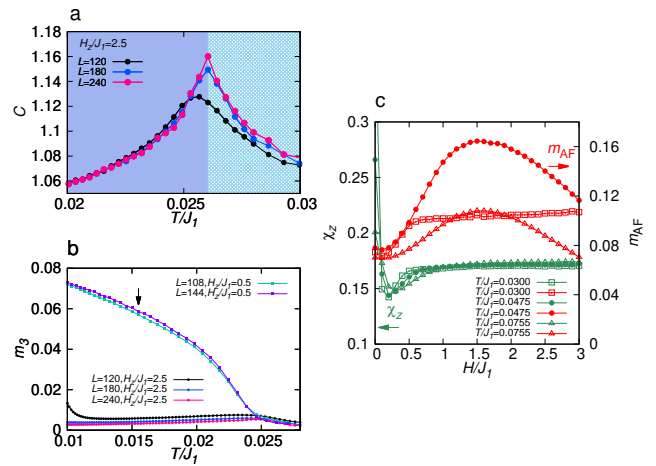}
 \caption{(a) The temperature and size dependence of the specific heat per spin near the paramagnetic(ring-liquid)-to-ripple transition temperature $T_c$ for $J_2/J_1=0.18$ and $H/J_1=2.5$, a magnified view of Fig.~\ref{Fig1}(e).
(b) The temperature dependence of the $m_3$ order parameter describing the lattice $C_3$ breaking for $J_2/J_1=0.18$ and $H/J_1=0.5$, 2.5. The arrow indicates the point at which the spin structure factor is shown in Fig.~S1(h) of Supplemental Material.
(c) The field dependence of the AF order parameter $m_{\rm AF}$ of the spin $xy$-component and the differential magnetic susceptibility ${\chi_z}$, at three representative temperatures in the paramagnetic regime for $J_2/J_1=0.18$. The lattice is a $L$=36 hexagonal cluster under open BC.}
 \label{Fig4}
\end{figure}

 In the present model, the AF short-range order (SRO) can be induced by applying magnetic fields to the ``pancake-liquid" or ``ring-liquid" states (see also Fig.~\ref{Fig1}(f)) \cite{Okumura}. We compute the field-dependence of the AF order parameter $m_{\rm AF}\equiv \frac{1}{2} |m_{A}^{xy}-m_{B}^{xy}|$, where $m_{A(B)}^{xy}$ is the transverse component of the sublattice magnetization per spin of the sublattice A(B), together with the differential magnetic susceptibility (slope of the magnetization curve) ${\chi_z}$. Fig.~\ref{Fig4}(c) shows the data at representative temperatures in the paramagnetic regime. The field turns out to enhance the AF SRO in the pancake (ring)-liquid region, its onset accompanied by the rise in $\chi_z$. Of course, our model is 2D, and this field-induced antiferromagnetism is only a SRO, in contrast to experiments on ${\rm Bi_3 Mn_4 O_{12} (NO_3)}$ \cite{Matsuda}. Non-negligible interlayer coupling in ${\rm Bi_3 Mn_4 O_{12} (NO_3)}$, both within a bilayer and between neighboring bilayers, may play a role in stabilizing the experimentally observed AF long-range order (LRO) and the magnetization jump.

 We wish to discuss possible experimental realizations of the ripple state. One candidate material might be the $S$=3/2 bilayer-honeycomb Heisenberg AF ${\rm Bi_3 Mn_4 O_{12} (NO_3)}$. It was reported that the material exhibited not only the disordered ground state \cite{Smirnova} but also the field-induced antiferromagnetism \cite{Matsuda}, which may suggest that the material might lie close to the AF phase boundary. Although the ripple \textcolor{black}{state} in the present model possesses only the quasi-LRO in the spin $xy$-components, the interlayer coupling inherent to real magnets would change it to the true LRO. Yet, the ripple state of essentially the same character as that in 2D would exist with the ripple-like spin LRO and the characteristic ring-like $S_{\perp}({\bf q})$ with unbroken $C_3$ symmetry. Further experiments on the in-field ordering of ${\rm Bi_3 Mn_4 O_{12} (NO_3)}$, preferably with a single crystal, would be desirable~\cite{comment}.

We finally comment that the ripple spin texture can induce a giant electric-polarization vortex via the spin current mechanism~\cite{Kimura, Goto, Yasui, Katsura_H}, which might be important for the design of nanoscale devices such as high-density memories and high-performance energy-harvesting devices~\cite{Naumov, Tang, Yadev}.(see also Supplemental Material)

In summary, by extensive MC simulations, we confirm the existence of a new type of multiple-$q$ state, ripple state, in the $J_1$-$J_2$ classical honeycomb-lattice Heisenberg antiferromagnet under magnetic fields. The spin texture there looks like a water ripple. The possible realization of the ripple state could be expected in the frustrated bilayer-honeycomb material ${\rm Bi_3 Mn_4 O_{12} (NO_3)}$ and the bilayer-kagome material ${\rm Ca_{10} Cr_7 O_{28}}$.

\begin{acknowledgments}
T.S. thanks Rico Pohle, Han Yan, Nic Shannon, Bella Lake, Johannes D. Reim, Karlo Penc, Romain Sibille, Shunsuke C Furuya and Hirohiko Shimada for fruitful discussions. This work is supported by the Theory of Quantum Matter Unit of the Okinawa Institute of Science and Technology Graduate University (OIST) and also supported by JSPS KAKENHI Grant Number 25247064, 17H06137 and 19K14665. Parts of the numerical calculations are performed using the facilities of the Supercomputing Center, ISSP, the University of Tokyo, and of OIST.
\end{acknowledgments}

%

\vspace{10pt}

\newpage
\clearpage

        \renewcommand{\thetable}{S\arabic{table}}%
        \setcounter{figure}{0}
        \renewcommand{\thefigure}{S\arabic{figure}}%
\renewcommand{\thesection}{\Roman{section}}
\makeatother
\setcounter{figure}{0} 
\setcounter{equation}{0}

\onecolumngrid
\begin{center} {\bf \large Ripple state in the frustrated honeycomb-lattice antiferromagnet \\
-Supplementary information-} \end{center}
\vspace{0.5cm}
\twocolumngrid

\author{Tokuro Shimokawa} 
\email{tokuro.shimokawa@oist.jp}
\affiliation{Okinawa Institute of Science and Technology Graduate University, Onna, Okinawa, 904-0495, Japan}
\author{Hikaru Kawamura}
\affiliation{Department of Earth and Space Science, Graduate School of Science, Osaka University, Toyonaka, Osaka 560-0043, Japan}

\date{\today}
\maketitle

\section{Numerical method}
Our Monte Carlo (MC) simulations are performed on the basis of the standard heat-bath method combined with the over-relaxation and the temperature-exchange methods. Unit MC step consists of one heat-bath sweep and 5-10 over-relaxation sweeps. To take account of the strong incommensurability effect, we treat mainly hexagonal finite-size clusters with a trigonal symmetry under open BC as shown in Fig.~1(a) of the main text. To investigate the spontaneous translational-symmetry breaking in the ripple state, we also treat diamond clusters under periodic BC. The hexagonal (diamond) cluster contains $N$=$3L^2/2 \ (2L^2)$ spins under open (periodic) BC with $36 \leq L \leq 240$, $N$ being the total number of spins. Typically, our MC runs contain $\sim 10^7$ MC steps, and averages are made over 1-3 independent runs. Note that, in calculating the real-space spin/chirality configuration and the spin structure factor, we cut the temperature-exchange process. After thermalizing the system, we monitor the symmetry-breaking pattern typically during $10^3 \sim 10^4$ MC steps to detect the real-space spin/chirality configurations, and $10^4 \sim 10^5$ MC steps to compute the static spin structure factor.

We define several physical quantities computed in our MC simulations. The $xy$-component of the sublattice spin structure factor is defined by
\begin{eqnarray}
S_{\perp}({\bf q})=\frac{2}{N}  \sum_{\mu=x,y}\langle |\sum_{j} S_j^{\mu} e^{i {\bf q} \cdot {\bf r}_j}|^2 \rangle,
\end{eqnarray}
where ${\bf r}_j$ is the position vector of the spin at site $j$ on one particular triangular sublattice, ${{\bf S}_j^{\mu}}$ is the spin at the site $j$ on the sublattice ($A$ or $B$), ${\bf q}=(q_x,q_y)$ is the associated wavevector,  $\langle \cdots \rangle$ is a thermal average, and the lattice constant is taken to be the NNN neighbor distance of the original honeycomb lattice. Recall that the honeycomb lattice consists of two interpenetrating triangular sublattices. Hence in our plots of $S_{\perp}({\bf q})$, the ${\bf q}=(0, 0)$ point corresponds to the AF point of the original honeycomb system.

 The local scalar chirality $\chi$ is defined for three neighboring spins located on each elementary triangle $t$, either upward or downward,  on either sublattice A or B, by
\begin{equation}
\chi_{t^{A(B)}} = \langle {\bf S}_{t^{A(B)1}} \cdot ({\bf S}_{t^{A(B)2}} \times {\bf S}_{t^{A(B)3}}) \rangle,
\end{equation}
where $t^{A(B)}$ denotes an elementary triangle on the sublattice $A(B)$, while ${\bf S}_{t^{A(B)1}} \sim {\bf S}_{t^{A(B)3}}$ are three Heisenberg spins on an elementary triangle numbered in a clockwise fashion on a triangle.

The $m_3$ order parameter describing the lattice $C_3$ symmetry breaking is defined by
\begin{eqnarray}
m_3=\langle |{\bf m}_3| \rangle, \ \  {\bf m}_3=\epsilon_1 {\bf e}_1 + \epsilon_2 {\bf e}_2 + \epsilon_3 {\bf e}_3,
\end{eqnarray}
where ${\bf e}_{1,2,3}$ are unit vectors running along the NN directions of the honeycomb lattice, {\it i.e.\/}, ${\bf e}_1=(0,1)$, ${\bf e}_2=(-\sqrt{3}/2,-1/2)$ and ${\bf e}_3=(\sqrt{3}/2,-1/2)$, and $\epsilon_{1,2,3}$ are the total NN bond energy normalized per bond along the ${\bf e}_{1,2,3}$ directions, respectively.

On the basis of the spin-current model, we define the local electric-polarization vector ${\bf P}_t^{A(B)}$ on each elementary triangle $t$, either upward or downward, on the sublattice A or B by 
\begin{eqnarray}
{\bf P}_{t^{A(B)}}=\langle \frac{1}{3} \sum_{i,j \in t^{A(B)}} {\bf e}_{i,j} \times ({\bf S}_i^{(xy)} \times {\bf S}_j^{(xy)}) \rangle,
\end{eqnarray}
where ${\bf S}_i^{(xy)}$ is an $xy$-component of the spin ${\bf S}_i$, and ${\bf e}_{i,j}$ is a unit vector from the site $i$ to the site $j$. There are three bonds on every elementary triangle and we take an average for the three bonds on each elementary triangle.

\section{Paramagnetic, spin-liquid and single-$q$ spiral states}
\begin{figure}[t]
  \includegraphics[bb=0 0 612 792, width=10.5cm,angle=0]{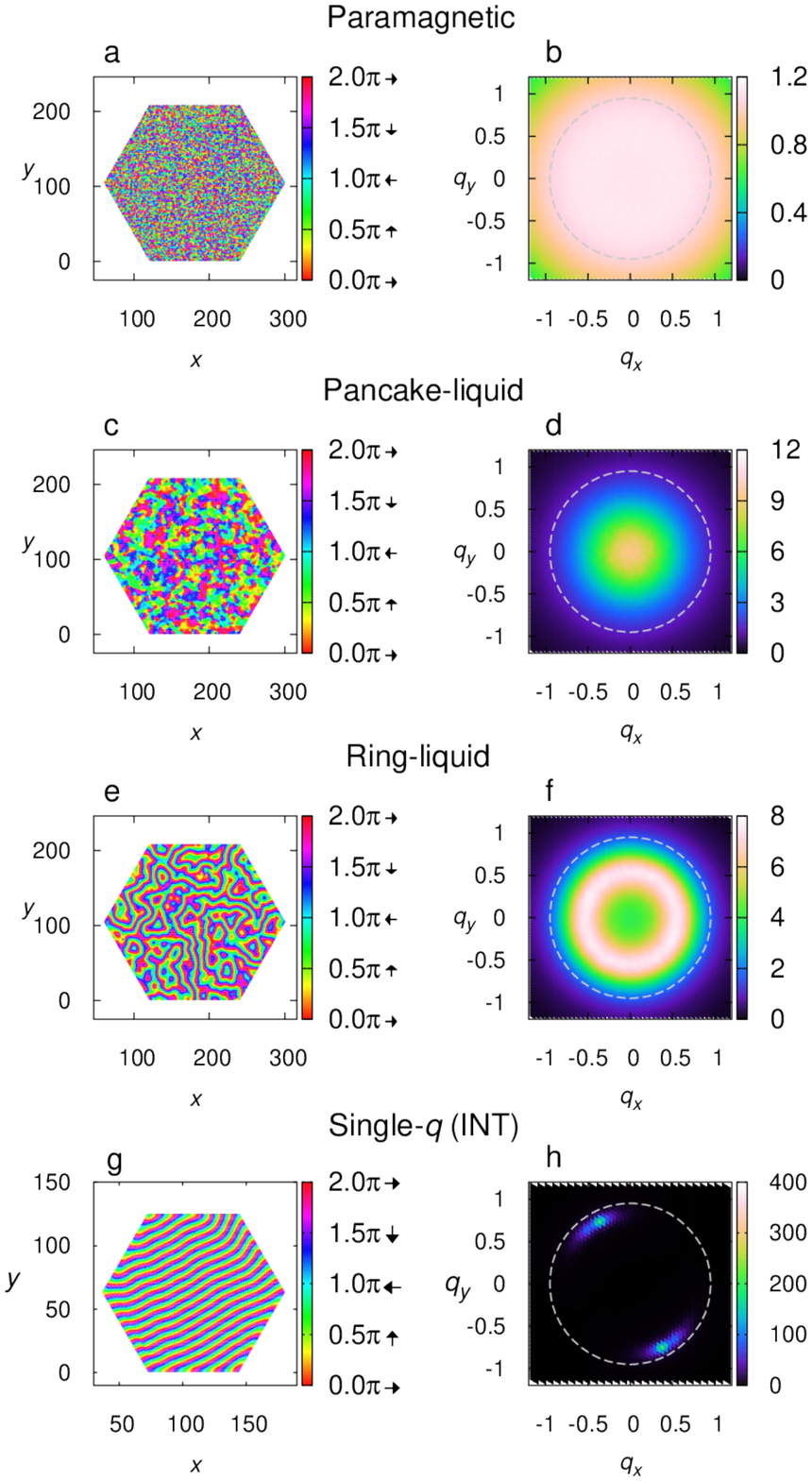}
 \caption{The real-space sublattice $xy$-spin configuration and the corresponding static spin structure factor in the paramagnetic and the single-$q$ states.
(a),(c),(e),(g), The real-space sublattice $xy$-spin configuration at $J_2/J_1=0.18$ for the high-temperature paramagnetic state (a), the pancake-liquid state (c), the ring-liquid state (e), and the single-$q$ spiral state (g). The color displays the angle of the $xy$ spin at each site. (b),(d),(f),(h), The intensity plot of the sublattice static spin structure factor $S_{\perp} ({\bf q})$ at $J_2/J_1=0.18$ for the high-temperature paramagnetic state (b), the pancake-liquid state (d), the ring-liquid state (f), and the single-$q$ helical state (h). The dashed gray line denotes the ground-state degeneracy line for $J_2/J_1$=0.18. The field intensity is $H/J_1=2.5$ (a)-(f) and 0.5 (g),(h) while the temperature is $T/J_1=0.14$ (a),(b), 0.06057 (c),(d), 0.028 (e),(f), and 0.014996 (g),(h). The lattice size is $L$=240 (a)-(f) and 144 (g),(h) under open BC. } 
 \label{Fig2}
\end{figure}

 We show in Fig.~\ref{Fig2}(a-f) the real-space sublattice $xy$-spin configuration and the associated sublattice spin structure factor $S_{\perp}({\bf q})$, each for the high-$T$ paramagnetic state (a, b), the pancake-liquid state (c, d), and the ring-liquid state (e, f). In the high-$T$ paramagnetic state, spins exhibit a `gas-like' behavior, the corresponding $S_{\perp}({\bf q})$ exhibiting a very diffuse pattern (Fig.~\ref{Fig2}(b)). By contrast, the pancake- and the ring-liquid states exhibit enhanced short-range spin correlations as shown in Fig.~\ref{Fig2}(c), (e).  As observed in Ref.~\cite{Okumura}, the corresponding $S_{\perp}({\bf q})$ show characteristic structures in the $q$-space shown in Fig.~\ref{Fig2}(d) and (f). In the ring-liquid state, a pronounced ring-like pattern arises, while, in the pancake-liquid state, the inside of the ring is `buried'. From a thermodynamic viewpoint, although these two states are still paramagnetic states, they may be regarded as cooperative paramagnets due to their high degrees of correlations developed in $q$-space, to be distinguished from the standard high-$T$ paramagnetic state.

 For the  single-$q$ (INT) state, the corresponding quantities are shown in  Fig.~{\ref{Fig2}(g)}, (h). As can be seen from Fig.~\ref{Fig2}(g), the sublattice $xy$-spin components form a spiral running along the direction intermediate between the NN and NNN directions, leading to a pair of single-$q$ wavevectors $\pm q^*$ chosen from the infinitely many $q$ states on the degenerate ring.

\section{Relation between the orders on the two sublattices in the ripple state}

\begin{figure}[t]
  \includegraphics[bb=160 100 800 500,width=12.0cm,angle=0]{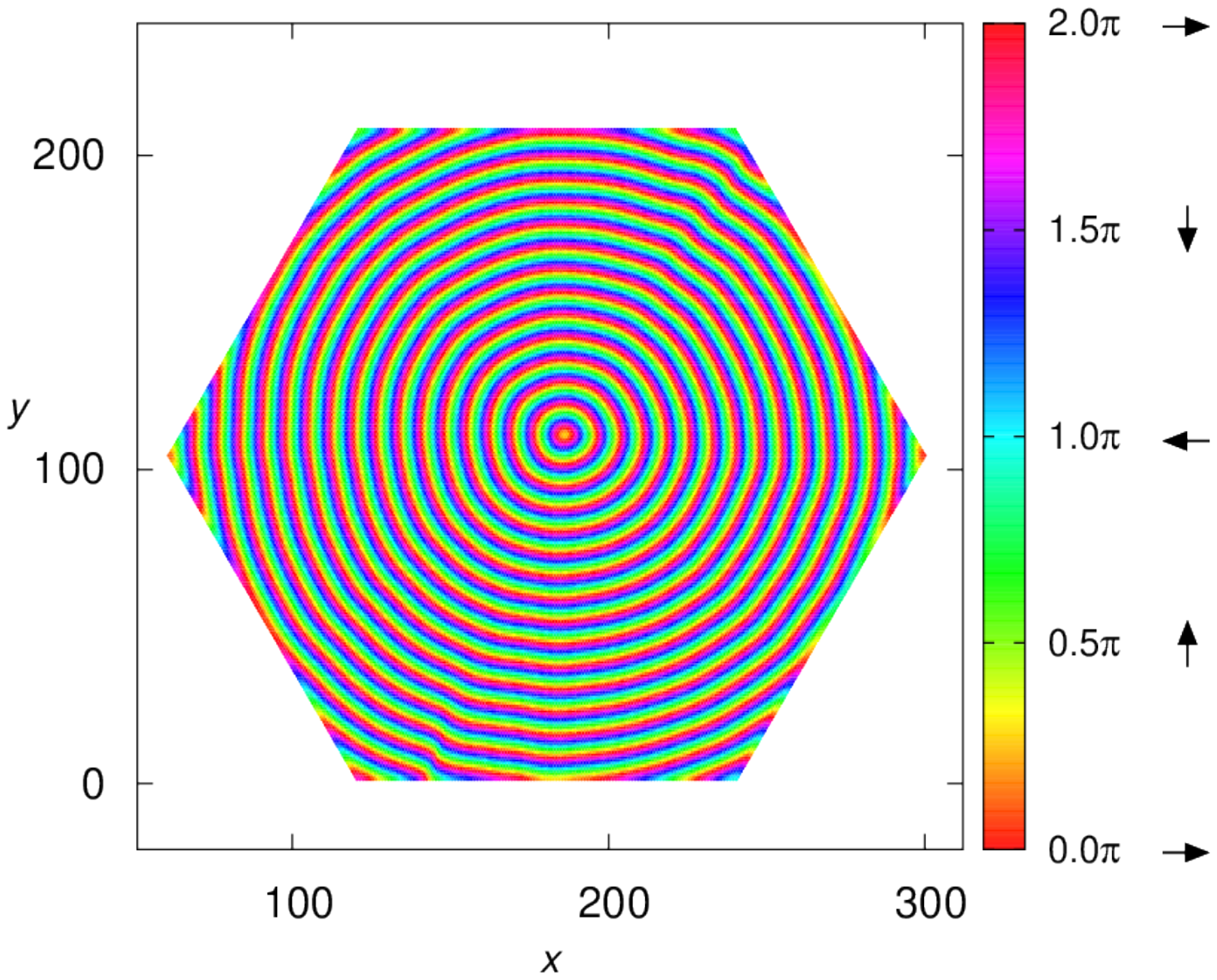}
 \caption{Real-space $xy$-spin configuration for the sublattice B.  The color plot of the direction of the $xy$-spin on the sublattice B in the ripple state at $J_2/J_1=0.18$, $T/J_1=0.01\textcolor{black}{0294}$ and $H/J_1=2.5$ for the system size $L$=240 ($N$=86400 spins) under open BC, which should be compared with the corresponding plot for the sublattice A shown in Fig.~1(c) of the main text.}
 \label{Sup1}
\end{figure}

 We comment on the interrelation between the spin orders on the two interpenetrating triangular sublattices A and B in the ripple state. In all types of ordered states observed in the present calculation, the phase (angle) difference $\alpha$ between the two NN spins on the two sublattices are nearly $\pi$. So, if one rotates the $xy$-spin on one sublattice by nearly $\pi$, one gets the $xy$-spin configuration on the other sublattice which might be related to the fact that $J_1$ interaction between two sublattices is antiferromagnetic. 
In Fig.~\ref{Sup1}, we show the real-space sublattice $xy$-spin configuration in the ripple state for the other sublattice (sublattice B) than that depicted in Fig.~1(c) of the main text for the sublattice A. The parameters are taken common as those of Fig.~1(c) of the main text, {\it i.e.\/}, $J_2/J_1=0.18$, $T/J_1=0.01$, $H/J_1=2.5$ and $L$=240 ($N$=86400 spins) under open BC. If one compares Fig.~\ref{Sup1} with Fig.~1(c) of the main text, one sees that essentially the same spin configurations are realized between the two sublattices except for a phase shift of nearly $\pi$.

\begin{figure*}[t]
  \includegraphics[bb=0 80 792 612,width=18.0cm,angle=0]{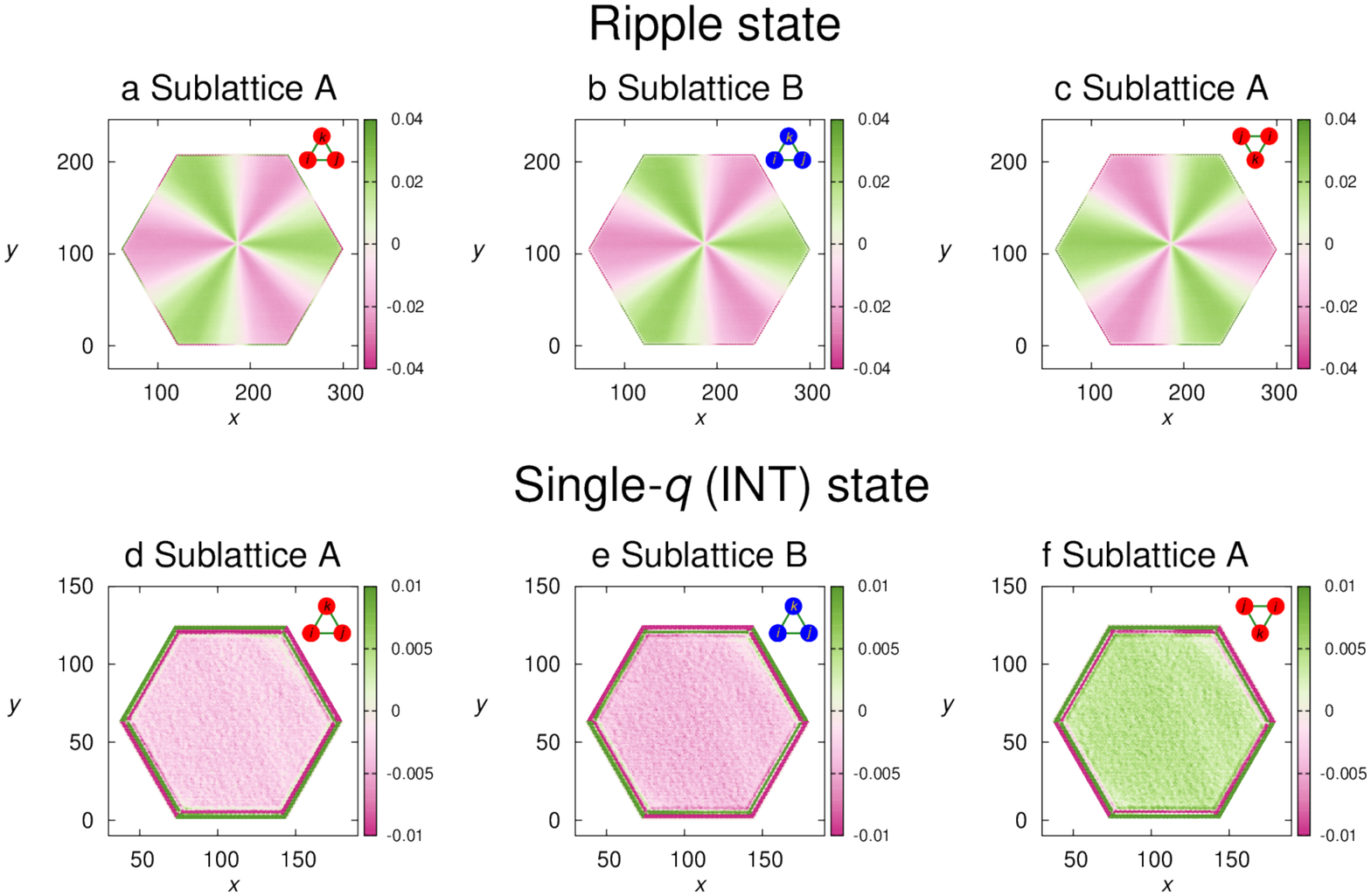}
 \caption{Real-space scalar-chirality patterns in the ripple and the single-$q$ states.
(a-c) The ripple state at $J_2/J_1=0.18, T/J_1=0.010294$ and $H/J_1=2.5$. The local scalar chiralities are defined on (a) upward triangles on the sublattice A, (b) upward triangles on the sublattice B, (c) downward triangles on the sublattice A. 
(d-f) The single-$q$ (INT) state at $J_2/J_1=0.18, T/J_1=0.01499$ and $H/J_1=0.5$. The local scalar chiralities are defined on (d) upward triangles on the sublattice A, (e) upward triangles on the sublattice B, (f) downward triangles on the sublattice A. }
 \label{Sup2}
\end{figure*}

\begin{figure*}[t]
  \includegraphics[bb=50 200 792 612, width=19cm,angle=0]{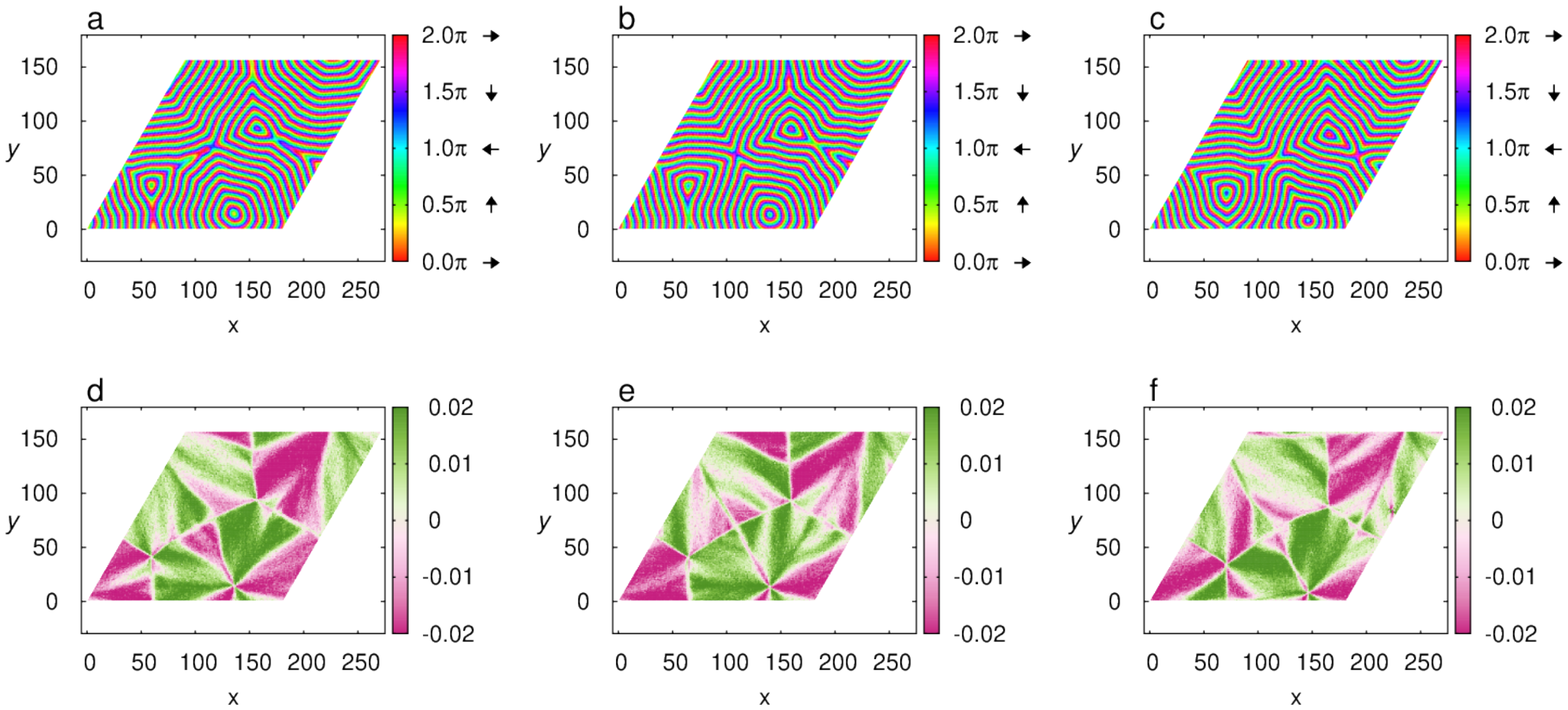}
 \caption{MC step dependence of the spin and the chirality patterns in the ripple state.
(a-c) The sublattice $xy$-spin configurations.
(d-f) The corresponding local scalar chirality configurations.
During its dynamical evolution, the systems stays at full equilibrium in the ripple state at $J_2/J_1=0.18$, $T/J_1=0.01$ and $H/J_1=2.5$. The lattice is the $L=180$ diamond cluster under periodic BC. After initial full thermalization, we wait for 0 (a), (d), $10^5$ (b), (e), and $2\times 10^5$ (c), (f), MC steps, and the shots are taken by the short-time average of 1000 MC steps.
}
 \label{Sup3}
\end{figure*}

\section{Real-space chirality patterns in the ripple and the spiral states}

Scalar chirality defined on each upward (or downward) elementary triangle on each sublattice (A or B) is also useful, especially in probing the $Z_2$-symmetry breaking. The spatial patterns of the scalar chirality computed for $J_2/J_1=0.18$ are shown in Fig.~\ref{Sup2} for the ripple state ($T/J_1=0.010294$, $H/J_1=2.5$) in the upper row, and for the single-$q$ spiral state ($T/J_1=0.01499$, $H/J_1=0.5$) in the lower row. The lattice is the hexagonal cluster under open BC of the size $L=240$ (upper row) and $L=144$ (lower row).

 In the ripple state, reflecting its ripple-like $xy$-spin configuration, the scalar chirality happens to vanish along the six symmetric lines emanating from the sample core, which corresponds to the NN direction of the original honeycomb lattice, and the chirality sign alternates in crossing these six emanating lines, forming apparent domain-like patterns. We note that such a chiral-domain-like pattern is an immediate consequence of the ripple-like spin configuration. In the single-$q$ spiral state, by contrast, the sign of the chirality is kept uniform over the entire sample. 

 As mentioned above, the chirality sign is common between the sublattices A and B both in the ripple and the single-$q$ spiral states, as long as one employs common upward (or common downward) triangles in the definition of the chirality. By contrast, the chirality sign is reversed when the chirality is defined on triangles of different orientations, upward vs. downward, as can be seen from Fig.~\ref{Sup2}(a),(c) and Fig.~\ref{Sup2}(d),(f).

 Since the sign of the scalar chirality tends to be reversed between the upward and the downward triangles, the total net chirality over the entire sample is expected to be cancelled out not only in the single-$q$ spiral states but also in the ripple state. Hence, no net topological Hall effect is expected, even in the apparent chiral domain of Fig.~\ref{Sup2}. This is in sharp contrast to the triple-$q$ skyrmion-lattice state having a total net scalar chirality \cite{Okubo}.

\section{Stability of the real-space spin/chirality configurations}

To examine the breaking of the translational, the U(1) and the $Z_2$ symmetries in the ripple state, we investigate the MC step dependence of the sublattice spin and the sublattice scalar-chirality configurations. The results obtained for the ripple state at $J_2/J_1=0.18$, $T/J_1=0.01$ and $H/J_1=2.5$ are shown in Fig.~\ref{Sup3}. The lattice is the $L=180$ diamond cluster under periodic BC. Periodic BC is chosen here to avoid the pinning effect due to surfaces, which is inevitable in open BC. Fig.~\ref{Sup3}(a-c) represent the sublattice $xy$-spin configuration, and Fig.~\ref{Sup3}(d-f) the corresponding local scalar-chirality configurations. After initial thermalization employing the heat-bath, the over-relaxation and the temperature-exchange methods, we cut the temperature-exchange process and wait for 0 (a,d), $10^5$ (b,e) and $2\times 10^5$ (c,f) MC steps, and the shots are taken by the short-time average of 1000 MC steps to reduce the thermal noise.

 As can be seen from these figures, even under periodic BC, the real-space spin and the chirality configurations are nearly locked during a long MC time of $2\times 10^5$ MC steps. This observation suggests that the translation, the $U(1)$ spin-rotation and the $Z_2$ symmetries are spontaneously broken in the ripple state. Of course, since our model is 2D, the breaking of the translation and the U(1) symmetries should be at most algebraic at nonzero temperatures, whereas the $Z_2$ symmetry could fully be broken. Such a symmetry-breaking pattern seems to be consistent with the observed second-order nature of the transition between the ripple and the ring-liquid states accompanied by the weakly divergent anomaly in the specific heat shown in Fig.~4(a) of the main text.

\section{The position of the ripple core}

\begin{figure}
 \includegraphics[bb=130 450 630 750,width=12.0cm,angle=0]{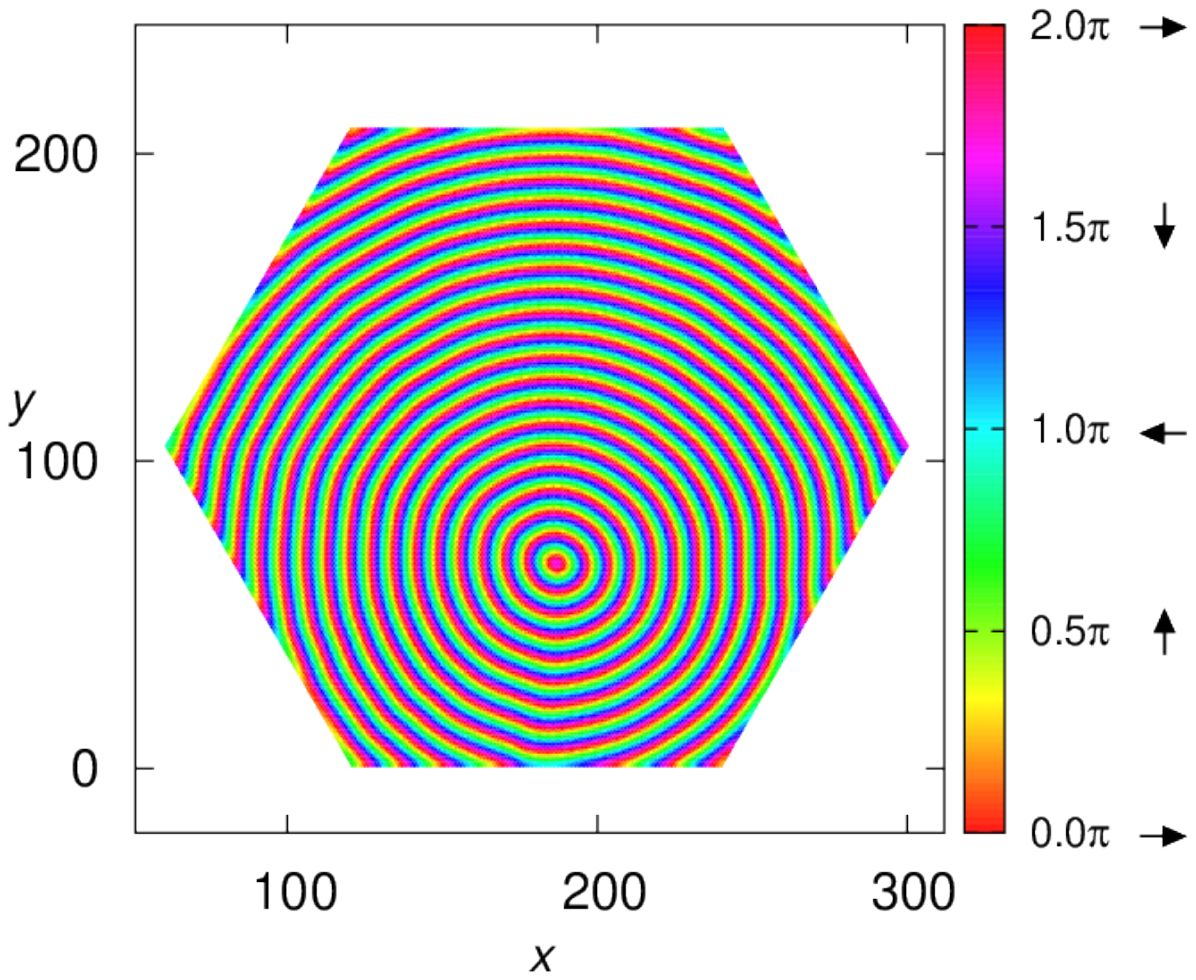}
 \caption{\textcolor{black}{Real-space sublattice $xy$-spin configuration in the ripple state at $J_2/J_1=0.18$ obtained for exactly the same parameter sets as those of Fig.~1(c) of the main text, {\it i.e.\/}, $H/J_1=2.5$, $T/J_1=0.01\textcolor{black}{0294}$ and $L$=240 under open boundary condition, with using different initial spin configurations and different random-number sequences.
}
}
 \label{Sup-bet}
\end{figure}

\textcolor{black}{Interestingly, the ripple core is not always located at the center of the lattice even under open boundary condition. An example of such ripple with its core position largely deviating from the center is shown in Fig.~\ref{Sup-bet}, which is obtained for exactly the same parameter sets as those of  Fig.~1(c) of the main text, {\it i.e.\/}, $H/J_1=2.5$, $T/J_1=0.010294$ and $L$=240 under open boundary condition, but with using different initial spin configurations and different random-number sequences. Essentially the same behavior is observed also for smaller lattices. While the ripple core is located sometimes near and sometimes off the center depending on each MC run, it turns out that the associated energy hardly differs from each other. In fact, the energies of different ripple patterns with different locations of the core agree with the high accuracy, say, with the relative accuracy of $10^{-7}$.  Reflecting the translation-symmetry breaking in the ripple state, the core position is almost locked in time during the MC run. Yet, concerning the core position on the lattice, there is no energetical preference. The ripple core can be anywhere on the lattice, not only for periodic BC but also for open BC.
}

\section{Phenomenological equation describing the ripple state}

We construct here a phenomenological equation reproducing the observed ripple-like spin configurations and the chiral domain patterns. We begin with the following sublattice $xy$-spin configurations in the single-$q$ spiral state,
\begin{eqnarray}
{\bf S}_{i}^{\mu} ({\bf r}_i) \propto  [{\rm cos} ({\bf q}_{} \cdot {\bf r}_i + \theta_{}+\alpha \delta_{\mu B}) {\bf e}_{1} \nonumber \\
 + \ \ Z {\rm sin} ({\bf q}_{} \cdot {\bf r}_i + \theta_{}+\alpha \delta_{\mu B}) {\bf e}_{2}],
\end{eqnarray}
where $\theta$ is an arbitrary phase factor, $\alpha$ is the phase difference between the two sublattices, ${\bf e}_{1}$ and ${\bf e}_{2}$ are mutually orthogonal unit vectors in the spin $xy$-plane (${\bf e}_{1}$ $\perp$ ${\bf e}_{2}$), $Z=\pm 1$ is an integer describing the $Z_2$ symmetry breaking, and $\delta_{\mu B}$ is a Kronecker delta. In the case of $J_2/J_1=0.18$ of our interest, $\alpha$ was reported to take a value close to $\pm \pi$ (see also \cite{Katsura, Shimokawa}). 

The ripple-like spin texture looks like to have single-$q$ spiral configuration along any direction from the ripple core. So, we expect that the ripple spin configuration can be constructed by an infinite sum of the single-$q$ spiral states composed by distinct ordering wavevectors lying on the degenerate ring. This simple idea leads to the equation,
\begin{eqnarray}
{\bf S}_{i}^{\mu} ({\bf r}_i) \propto  \sum_{m=1}^{N_q} [\ {\rm cos} ({\bf q}_{m}\cdot {\bf r}_i + \theta_{m}+\alpha \delta_{\mu B}) {\bf e}_{m,1} \nonumber \\
 + \ \ Z {\rm sin} ({\bf q}_{m}\cdot{\bf r}_i + \theta_{m}+\alpha \delta_{\mu B}) {\bf e}_{m,2} ],
\end{eqnarray}
where ${\bf q}_m$ are wavevectors on the degenerate ring, $m$ means an index of each wavevector, $\theta_{m}\equiv \frac{m\pi}{N_q}$, and $N_q$ is the number of treated wavevectors. We find, however, that this simple equation cannot well reproduce the observed ripple-like spin texture.

\begin{figure}
  \includegraphics[bb=100 0 560 792,width=12.0cm,angle=0]{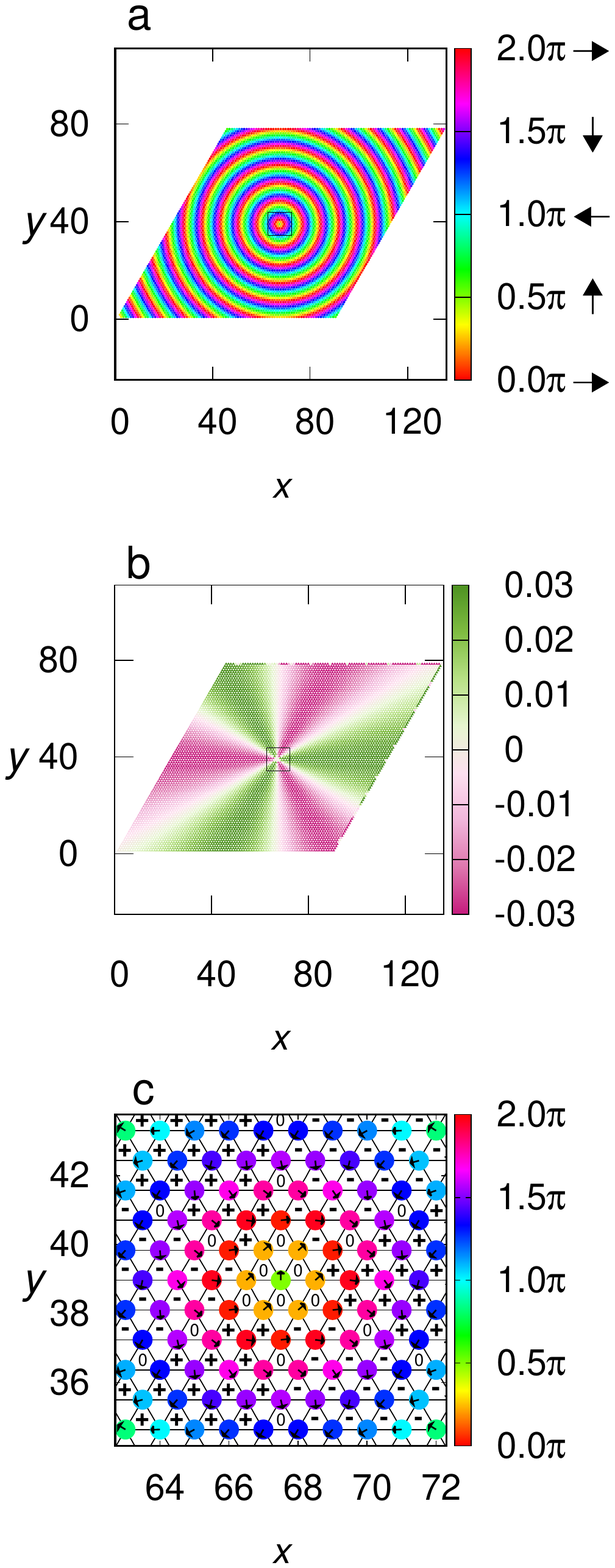}
 \caption{Spin and chirality configurations obtained from the ripple equation.
(a) The real-space sublattice $xy$-spin configuration for $J_2/J_1=0.18$. 
(b) The corresponding local scalar chirality configuration.
(c) The focused view around the ripple core indicated by the boxes in (a) and (b), whereas $+$,$-$ and $0$ shows the sign of the local scalar chirality on each upward elementary triangle.}
 \label{Sup4}
\end{figure}

The equation describing a real water ripple is known to have a form like $\sim \frac{1}{qr} {\rm sin}(q r)$ ($q \equiv |{\bf q}|$, $r \equiv |{\bf r}|$). In view of this, we modify the above eq.~(6) by replacing ${\bf q}_{m} \cdot {\bf r}_i$ by $|{\bf q}_{m}| |{\bf r}_i|$, and obtain the following equation,
\begin{eqnarray}
{\bf S}_{i}^{\mu} ({\bf r}_i) \propto  \sum_{m=1}^{N_q} [\ {\rm cos} (|{\bf q}_{m}| |{\bf r}_i| + \theta_{m}+\alpha \delta_{\mu B}) {\bf e}_{m,1} \nonumber \\
  \ \ \pm {\rm sin} (|{\bf q}_{m}| |{\bf r}_i| + \theta_{m}+\alpha \delta_{\mu B}) {\bf e}_{m,2} ].
\end{eqnarray}
where we put $Z=\pm 1$. 

Fig.~\ref{Sup4}(a) and (b) shows the $xy$-spin and the scalar-chirality configurations obtained by this modified equation (7), where we put
\begin{equation}
\left\{
\begin{array}{lll}
{\bf e}_{m,1}&\equiv&({\rm cos}2\theta_m , {\rm sin}2\theta_m ) \\
{\bf e}_{m,2}&\equiv&(-{\rm sin}2\theta_m , {\rm cos}2\theta_m ) , \\
\end{array}
\right.
\end{equation}
and we take $N_q=360$ wavevectors from the degenerate ring in $S_{\perp}({\bf q})$ observed in our MC simulation. As can be seen from the figure, the results can well reproduce the MC results shown in Fig.~1(c) of the main text and ~\ref{Sup2}(a).

The Fig.~\ref{Sup4}(c) shows the focused view around the ripple core in Fig.~\ref{Sup4}(a). The sign of the local scalar chirality is also given by $+$, $-$ or $0$ symbol in each elementary triangle. One notices that the local scalar chirality vanishes on any elementary triangle located along the NNN direction of the triangular sublattice (the NN direction on the original honeycomb lattice). This is simply because two out of three spins on any elementary triangle along the NNN direction of the triangular sublattice take a common direction under eq.~(7), leading to a zero scalar chirality. Hence, the apparent chiral domain structure with the $C_3$ symmetry comes from the fact that the $xy$-spin orientation is determined solely by the distance from the ripple core.

 It should be noticed that the degenerate ring in the wavevector space is not a true circle. For example, a careful inspection of Fig.~1(d) of the main text reveals that the distance from the AF point to the ring is slightly different depending on the direction, though there is the $C_3$ rotation symmetry. As $J_2/J_1$ approaches the critical value $1/6$, the ring tends to be perfect, converging to the AF point. 

 If we assume the degenerate ring to be perfect, we can simplify the above equation (7) in the large-$N_q$ limit,
\begin{eqnarray}
{\bf S}_{i}^{\mu} ({\bf r}_i) &\propto&  \int_{0}^{2\pi} [ {\rm cos} (qr_i + \frac{\theta}{2}+\alpha \delta_{\mu B}) {\bf e'}_{\theta ,1} \nonumber \\
                      & & \ \ \ \ \ \ \ \pm {\rm sin} (qr_i + \frac{\theta}{2}+\alpha \delta_{\mu B}) {\bf e'}_{\theta ,2} ] d\theta \\
\rightarrow  
{\bf S}_{i}^{\mu} ({\bf r}_i) &\propto&  
   \left(
    \begin{array}{cc}
     \mp  {\rm sin}(qr_i +\alpha \delta_{\mu B} )   \\
       \pm {\rm cos}(qr_i +\alpha \delta_{\mu B})    \\
    \end{array}
  \right),
\end{eqnarray} 
where 
\begin{eqnarray}
{\bf e'}_{\theta,1}=({\rm cos}\theta , {\rm sin}\theta ) \nonumber \\
{\bf e'}_{\theta,2}=(-{\rm sin}\theta , {\rm cos}\theta ) \nonumber.
\end{eqnarray}
Under the assumption of the perfect ring, the ripple state could be considered as a ``spherical wave" similar to the real water ripple, in contrast to the single-$q$ spiral state which can be considered as a ``plane wave". To reproduce the observed non-perfect ring behavior in the static spin structure, however, we need to treat large number of ordering wavevectors having mutually different absolute values of $q$. In this sense, one may consider the ripple state as an ${\it infinituple}$-$q$ state.

\section{Polarization vortex via spin-current mechanism}
\begin{figure}[t]
  \includegraphics[bb=50 90 612 792, width=10.0cm,angle=0]{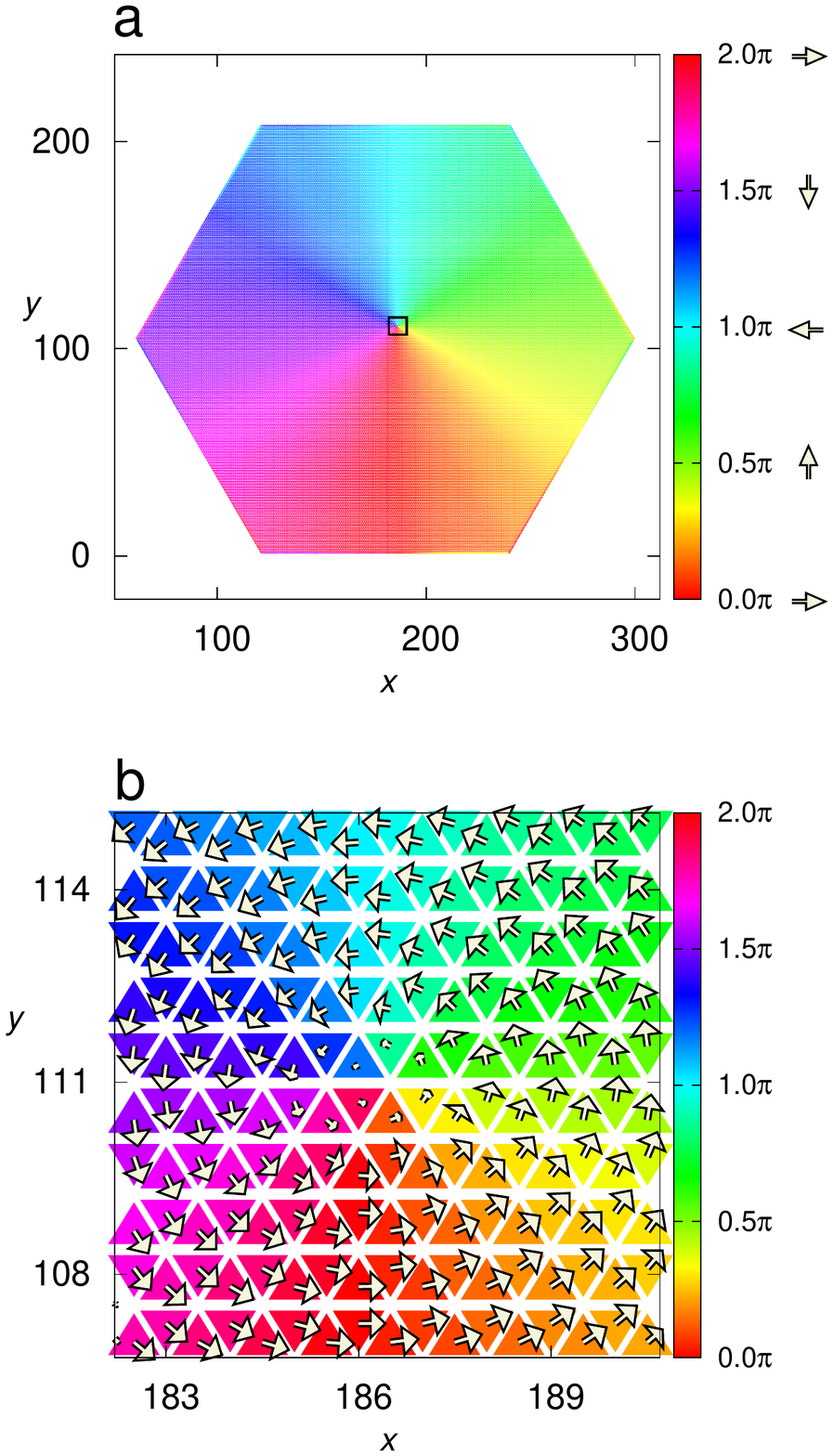}
 \caption{(a) The real-space electric-polarization-vector configuration on a sublattice induced via the spin-current mechanism. The corresponding sublattice $xy$-spin configuration is shown in Fig.~1(c) in the main text. The detailed definition of the local polarization vector is given by Eq. (4) of Supplemental Material. The circulations or the toroidal moments of the polarization vortices on the two sublattices are common in sign, meaning that the polarization vortex survives in the entire honeycomb-lattice system. The color means the direction in the $xy$-plane of the local polarization vector on each elementary triangle. (b) Focused view of the region denoted by the black square in (a). White arrows indicate the polarization vectors on each elementary triangle.}
 \label{Fig5}
\end{figure}

In this section, we wish to refer to a future potential application of the ripple state. Multiferroic properties of magnets, especially the ferroelectric orders induced by magnetic orders, have attracted much attention. Well-known examples might be frustrated magnets ${\rm TbMnO_3}$ \cite{Kimura}, ${\rm DyMnO_3}$ \cite{Goto} and ${\rm LiVCuO_4}$ \cite{Yasui}. In these materials, the vector chirality arising from the frustration-induced noncollinear spin order induces the spin-induced electric polarization via the so-called spin-current mechanism \cite{Katsura_H}. If the spin-current mechanism operates in the ripple spin texture, a giant vortex formed by electric-polarization vectors appears, as shown in Fig.~\ref{Fig5}. In the state, the vector chirality ${\bf \kappa}$ points either along $z/(-z)$ inducing the electric-polarization vortex in the $xy$ plane, which corresponds to the polarization vortex of $+/-$ circulation or $+/-$ toroidal moment, [${\bf r}\times {\bf P}({\bf r})$]$_z$, where ${\bf P}({\bf r})$ is the local polarization vector at the position ${\bf r}$. These two types of vortices with mutually opposite circulations become possible due to the symmetric nature of the interaction: If the interaction is antisymmetric as in  the Dzyaloshinskii-Moriya interaction, only one type of polarization vortex is possible. We note that such a giant electric-polarization vortex has been discussed in view of its tremendous potential for the design of nanoscale devices such as high-density memories and high-performance energy-harvesting devices \cite{Naumov, Tang, Yadev}. We hope that our present work provides a new route to create spin-induced electric-polarization vortex and stimulates future studies on development of nanoscale multiferroic devices.

\begin{thebibliography}{99}
\bibitem{comment}  \textcolor{black}{Another possible candidate might be a quantum spin liquid material, $S=1/2$ quantum bilayer-kagome magnet ${\rm Ca_{10} Cr_7 O_{28}}$~\cite{Balz, Balz2}. Very recently, it was suggested that this quantum material can be modeled as a semi-classical honeycomb-lattice Heisenberg model with the ferromagnetic (FM) NN and the AF NNN couplings, exhibiting a ring-liquid-like behavior with a characteristic ring-like pattern in the neutron scattering signal \cite{Biswas, Rico}. Although $J_1$ is FM in this material different from the one studied here, the ring-like degeneracy in the ground state still persists even in the $J_1<0$ model for $1/6<J_2/|J_1|<0.5$ \cite{Rastelli,Katsura}, and some of the ordering features may be common. Further study is desirable to clarify the situation.}
\end{thebibliography}
 \end{document}